\documentclass[onecolumn,english]{sig-alternate}
\usepackage[T1]{fontenc}
\usepackage[latin9]{inputenc}
\usepackage{babel}
\usepackage{array}
\usepackage{rotating}
\usepackage{url}
\usepackage{multirow}
\usepackage{amssymb}
\usepackage{graphicx}
\usepackage[unicode=true,
 bookmarks=false,
 breaklinks=true,pdfborder={0 0 1},backref=false,colorlinks=false]
 {hyperref}
\hypersetup{
 pdfauthor={Oscar Hernán Paruma-Pabón},
 pdfstartview=XYZ}
\usepackage{breakurl}

\makeatletter

\providecommand{\tabularnewline}{\\}

\usepackage[justification=centering]{caption}
\usepackage[para]{threeparttable}

\makeatother

\begin{document}
\CopyrightYear{2016}
\setcopyright{acmcopyright}
\conferenceinfo{CHASE'16,}{May 16 2016, Austin, TX, USA}
\isbn{978-1-4503-4155-4/16/05}\acmPrice{\$15.00}
\doi{http://dx.doi.org/10.1145/2897586.2897611}

\title{Finding Relationships between Socio-technical Aspects and Personality Traits by Mining Developer E-mails}

\author{\large Oscar Hernán Paruma-Pabón \textsuperscript{1} \\
\large ohparumap@unal.edu.co\\
\and
\\
\hspace{3cm} \\
\and
\large Fabio A. González \textsuperscript{1} \\
\large fagonzalezo@unal.edu.co\\
\and
\\
\large Jairo Aponte \textsuperscript{1} \\
\large jhapontem@unal.edu.co\\
\and
\\
\large Jorge E. Camargo \textsuperscript{2} \\
\large jorgecamargo@uan.edu.co\\
\and
\\
\large Felipe Restrepo-Calle \textsuperscript{1} \\
\large ferestrepoca@unal.edu.co\\
\and
\\
\large \textsuperscript{1} MindLab research group, Universidad Nacional de Colombia, Bogotá, Colombia
\\
\\
\large \textsuperscript{2} Lacser research group, Universidad Antonio Nariño, Bogotá, Colombia}
\maketitle
\begin{abstract}
Personality traits influence most, if not all, of the human activities,
from those as natural as the way people walk, talk, dress and write
to those most complex as the way they interact with others. Most importantly,
personality influences the way people make decisions including, in
the case of developers, the criteria they consider when selecting
a software project they want to participate. Most of the works that
study the influence of social, technical and human factors in software
development projects have been focused on the impact of communications
in software quality. For instance, on identifying predictors to detect
files that may contain bugs before releasing an enhanced version of
a software product. Only a few of these works focus on the analysis
of personality traits of developers with commit permissions (committers)
in Free/Libre and Open-Source Software projects and their relationship
with the software artifacts they interact with. This paper presents
an approach, based on the automatic recognition of personality traits
from e-mails sent by committers in FLOSS projects, to uncover relationships
between the social and technical aspects that occur during the software
development process. Our experimental results suggest the existence
of some relationships among personality traits projected by the committers
through their e-mails and the social (communication) and technical
activities they undertake. This work is a preliminary study aimed
at supporting the setting up of efficient work teams in software development
projects based on an appropriate mix of stakeholders taking into account
their personality traits.

\end{abstract}

\keywords{socio-technical aspects, FLOSS, personality traits, source code repository,
mailing lists, email communication.}

\section{Introduction}

The main motivation of this paper is to study the relationships between
software artifacts, mainly source code, and developers\textquoteright{}
personality traits to gain insight into the factors influencing developers
to write code for any or some modules in a FLOSS project. It is expected
these insights may help to explain the implicit mechanisms that lead
to self-forming software teams, and how the developers\textquoteright{}
personality marks are reflected in some technical actions such as
frequency of the commits, size of the sent patches, and terms used
in commit messages.

In the present work, we are looking for relationships between social
and technical aspects in the evolution of FLOSS projects, seeking
to answer the following research questions:
\begin{itemize}
\item what personality traits can be identified through communications among
software developers involved in FLOSS projects?,
\item what personality traits stand out according to the projects the software
developers are involved in?, and
\item what relationships can be observed between the social activities (communication
through the project mailing lists) of the committers and personality
traits characterizing the groups they belong to?
\end{itemize}
To the best of our knowledge, this paper is one of the first analyzing
personality traits of software developers in FLOSS projects and their
relationships with social and technical aspects.

The paper is structured as follows: In Section 2 we present a review
of the related work; in Section 3 we describe the methodology of the
study; Section 4 describes the experiments conducted and the results
obtained; Section 5 presents the conclusions, and lastly, we discuss
threats to validity in Section 6.

\section{Related work}

Some authors agree that most of the research into the software development
process has been focused on technical aspects \cite{key-4,key-5}.
However, some studies have addressed the role of human factors, e.g.
personality traits, in software engineering and software development.
Sheppard and Curtis\cite{key-2} report results from experiments to
determine the influence of human factors in software development.

Basili and Reiter\cite{key-3} remark that factors directly related
to the psychological nature of human beings play a major role in software
development. They concluded that research into the effects of human
factors on software is dependent on suitable measurement of several
non-functional software features. They report findings indicating
that a larger programming team size and the use of a disciplined methodology
have beneficial effects on the development process and the developed
product. 

In their review of productivity factors in software development, Wagner
and Ruhe\cite{key-4} give a special consideration of human factors
in software engineering. Such factors, as they explain, are often
not analyzed with equal detail as more technical factors further than
more than a third of the time a software developer is concerned with
other kind of work, not just technical work. One of the main contribution
of Wagner and Ruhe's work is the list of soft and technical factors
influencing productivity in software development they provided.

The work of Sommerville and Rodden\cite{key-5} discusses human, social,
and organizational factors affecting software processes and, to remark,
they discuss how to analyze software processes as human rather than
technical processes.

With regard to software process enhancements, Acuña and Juristo\cite{key-13}
proposed a Capabilities-oriented Software Process Model for assigning
people to roles according to their capabilities and the capabilities
demanded by the role, and empirically validated its positive impact
in software development effectiveness and efficiency. Along the same
line of work, seeking to associate personality with the software process,
Bradley and Hebert\cite{key-14} proposed a model that can be used
to analyze the personality type composition of an information system
development team and highlighted the impact of personality type on
team productivity. Capretz\cite{key-17} provides a personality profile
of software engineers according to the Myers\textendash Briggs Type
Indicator and the results of his study suggests that software engineers
are most likely to be ST (Sensing and Thinking) or TJ (Thinking and
Judgment) or NT (Intuition and Thinking). Most recently, Capretz and
Ahmed\cite{key-16} used the Myers-Briggs Type Indicator (MBTI), a
self-inventory designed to identify an individual personality type,
strengths, and preferences, to mapping job and skills requirements
to personality types for each of the activities involved in software
engineering processes such as system analysis, software design, programming,
testing, and maintenance. MBTI has also been part of the research
done by DaCunha and Greathead\cite{key-18}, Greathead\cite{key-19}.

Relying on one of the most widely used models of personality, Buchanan\cite{key-15}
explores the impact of the Big Five personality patterns on group
cohesiveness and group performance on creative tasks and establishes
patterns of three Big Five traits (Extraversion, Openness to Experience
and Conscientiousness) as potential predictors of group performance
on creative tasks. Kanij et al. \cite{key-20} based their work on
the question of \textquotedblleft whether the personality of software
testers may be different to other people involved in software development?\textquotedblright{}
and to test this hypothesis they collected personality profiles using
the Big Five factor model of a large group of software testers and
a large group of people involved in other roles of software development.
Their results indicate that software testers present a significantly
higher conscientiousness factor than other software development practitioners.

Although neither MBTI%
\footnote{{\tiny{}\url{en.wikipedia.org/wiki/Myers-Briggs_Type_Indicator\#Criticism}}%
} nor the Big Five are considered by all psychologists to be universally
accepted \cite{key-21}, many researchers are employing them for a
variety of purposes \cite{key-22}.

Studies such as those conducted by Yarkoni\cite{key-10}, Golbeck
et al.\cite{key-11} and Gill\cite{key-12} have sought to identify
personality traits from text (blogs, twitter, email). As referred
by Gill\cite{key-12}, \textquotedblleft Personality is projected
linguistically\textquotedblright{} and \textquotedblleft Personality
can be perceived through language\textquotedblright . The way people
write and speak and the words they use relate to their personality
traits, so one can say there is a strong relationship between personality
and the use of language, especially when people write or talk about
topics of their choice\cite{key-10}.

\section{Methodology}

\subsection{Datasets}

Building on the work done by Gonzalez-Barahona et al.\cite{key-6}
we used the data of the Eclipse project%
\footnote{{\tiny{}\url{www.eclipse.org/eclipse}}%
} available at %
\footnote{{\tiny{}\url{gsyc.es/~jgb/repro/2015-msr-grimoire-data}}%
}, with information from the following repositories: source code management
(git), issue tracking (Bugzilla), mailing lists (archived in mbox
format), and code review (Gerrit). From the dumps that are provided
by Metrics Grimoire, the databases were restored and the datasets
used in the experimental stage were built.

Since we are interested in identifying relationships between social
and technical aspects in the evolution of FLOSS projects, the source
code repository and the mailing lists are the most relevant data for
the purpose of this work. Specifically, we used the data of the Eclipse
Platform subproject, which in turn is divided into the following components
\cite{key-7}: Ant - Eclipse/Ant integration, Workspace (Team, CVS,
Compare, Resources) - Platform resource management, Debug - Generic
execution debug framework, Releng - Release Engineering, Search -
Integrated search facility, SWT - Standard Widget Toolkit, Text -
Text editor framework and UI - Platform user interface, runtime and
help components.

\subsection{Socio-Technical Analysis Methodology}

Because of the specificity of the study we conducted, it was necessary
to define a methodology to study socio-technical relationships in
FLOSS projects. The methodology we propose starts by defining the
best representation of the data describing the social and technical
aspects of the developers in the software development process to,
thereafter, build the datasets to be used in the experimental stage.
The representation we used for technical data was binary vectors.
Each vector represents whether a committer touched each file of the
project or not. For personality data, the representation was the personality
traits characterizing software developers, which they project through
their emails. An exploratory analysis was performed to become familiar
with the data and to identify potential inconsistencies that should
be corrected.

For each research question we wanted to answer, a specific experiment
was configured and carried out. The first experiment was intended
to answer RQ1: What personality traits can be identified through communications
between software developers involved in FLOSS projects?, so that,
at this stage, IBM Watson Personality Insights becomes more prominent.
The dataset consisted of emails sent by committers to the mailing
lists of the Eclipse Platform project and their subprojects.

The goal of the second experiment was to answer RQ2: What personality
traits stand out according to the projects the software developers
are involved in? and, at this stage, we used the personality traits
identified in the above stage, and using clustering techniques (k-means
and spectral clustering), we identified the personality traits characterizing
each of the resulting clusters.

Finally, the third experiment was intended to answer RQ3: What relationships
can be observed between the social activities (communication through
the project mailing lists) of the committers and personality traits
characterizing the groups they belong to? For this purpose we created
a graph (a social network) representing e-mail communication among
committers. Using the results obtained in the above stages, we determined
the more distinctive personality traits of the nodes connected to
the hubs in the graph.

\begin{figure}[h]
\begin{centering}
\includegraphics[width=1\columnwidth]{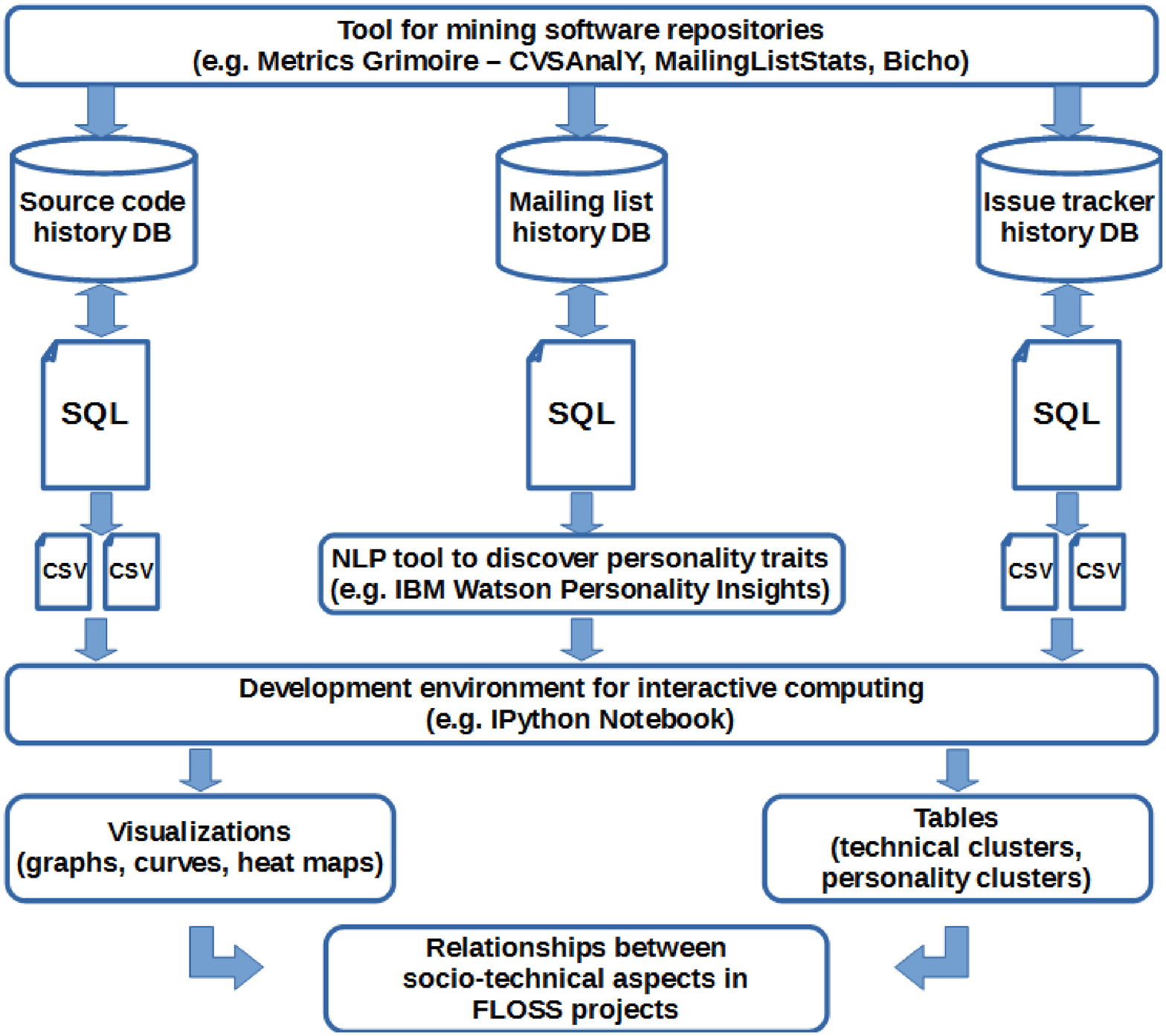}
\par\end{centering}

\centering{}\protect\caption{{\scriptsize{}Socio-Technical analysis methodology diagram.}}
\label{fig:socioTechnicalAnalysisMethodologyDiagram}
\end{figure}

The methodology we proposed is depicted in Figure \ref{fig:socioTechnicalAnalysisMethodologyDiagram}.
The main steps, which are described in detail in the following section,
were as follows:
\begin{itemize}
\item Restoring databases from the dumps (source code repository and mailing
lists).
\item Datasets construction (social, technical and personality).
\item Exploratory data analysis.
\item Identifying technical and personality groups by applying clustering
techniques.
\item Identifying personality traits that characterize each of the technical
groups.
\item Visualization of social (communication) networks.
\item Identification of social and technical relationships.
\end{itemize}

\subsection{Tools}

For clustering we used scikit-learn%
\footnote{{\tiny{}\url{scikit-learn.org}}%
}, for plotting we used matplotlib%
\footnote{{\tiny{}\url{matplotlib.org}}%
}, for scientific computing we used NumPy%
\footnote{{\tiny{}\url{www.numpy.org}}%
} and SciPy%
\footnote{{\tiny{}\url{www.scipy.org}}%
}; for data manipulation we used pandas%
\footnote{{\tiny{}\url{pandas.pydata.org}}%
}; and for network visualization we used NetworkX%
\footnote{{\tiny{}\url{networkx.github.io}}%
}.

On the other hand, with regard to the study of social aspects identified
from communications among software developers, the tool we used was
IBM Watson Personality Insights%
\footnote{{\tiny{}\url{www.ibm.com/smarterplanet/us/en/ibmwatson/developercloud/personality-insights.html}}%
}. IBM Watson Personality Insights service \cite{key-10,key-23} %
\footnote{{\tiny{}\url{www.ibm.com/smarterplanet/us/en/ibmwatson/developercloud/doc/personality-insights/science.shtml}}%
} can detect personality traits reflected in text written by a subject.
This was particularly useful for this work since it was unfeasible
to apply a personality test to each of the committers who contribute
to the FLOSS project under study.

\section{Experiments and Results}

\subsection{Exploratory data analysis}

To learn about the data to be used in the experiments, we conducted
an exploratory analysis summarized in Table \ref{tab:EclipsePlatformProjectNumberRegisters}.
The date range for which data were obtained is between January 1st,
2003 and January 1st, 2015.

\begin{table}[h]
\begin{centering}
\begin{tabular}{|>{\centering}m{0.5cm}|>{\centering}m{2.1cm}|>{\centering}m{0.9cm}|>{\centering}m{0.9cm}|>{\centering}m{0.9cm}|>{\centering}m{0.9cm}|}
\hline 
\textbf{\scriptsize{}No} & \textbf{\scriptsize{}Project / Subproject} & \begin{turn}{90}
\textbf{\scriptsize{}Committers}
\end{turn} & \begin{turn}{90}
\textbf{\scriptsize{}Commits}
\end{turn} & \begin{turn}{90}
\textbf{\scriptsize{}{~\parbox{1.7cm}{%
Mailing List Senders
}~}}
\end{turn} & \begin{turn}{90}
\textbf{\scriptsize{}{~\parbox{1.7cm}{%
Mailing List Messages
}~}}
\end{turn}\tabularnewline
\hline 
\hline 
\multirow{1}{0.5cm}{{\scriptsize{}1}} & \multirow{1}{2.1cm}{{\scriptsize{}Eclipse Platform}} & {\scriptsize{}46} & {\scriptsize{}6829} & {\scriptsize{}405} & {\scriptsize{}939}\tabularnewline
\hline 
\multirow{1}{0.5cm}{{\scriptsize{}2}} & \multirow{1}{2.1cm}{{\scriptsize{}Platform Text}} & {\scriptsize{}33} & {\scriptsize{}5911} & {\scriptsize{}71} & {\scriptsize{}454}\tabularnewline
\hline 
\multirow{1}{0.5cm}{{\scriptsize{}3}} & \multirow{1}{2.1cm}{{\scriptsize{}Platform UI}} & {\scriptsize{}112} & {\scriptsize{}25110} & {\scriptsize{}375} & {\scriptsize{}5069}\tabularnewline
\hline 
\multirow{1}{0.5cm}{{\scriptsize{}4}} & \multirow{1}{2.1cm}{{\scriptsize{}RelEng}} & {\scriptsize{}4} & {\scriptsize{}205} & {\scriptsize{}232} & {\scriptsize{}22716}\tabularnewline
\hline 
\multirow{1}{0.5cm}{{\scriptsize{}5}} & \multirow{1}{2.1cm}{{\scriptsize{}Resources}} & {\scriptsize{}28} & {\scriptsize{}3077} & {\scriptsize{}180} & {\scriptsize{}1561}\tabularnewline
\hline 
\multirow{1}{0.5cm}{{\scriptsize{}6}} & \multirow{1}{2.1cm}{{\scriptsize{}SWT}} & {\scriptsize{}46} & {\scriptsize{}21984} & {\scriptsize{}1125} & {\scriptsize{}5967}\tabularnewline
\hline 
\end{tabular}
\par\end{centering}

\centering{}\protect\caption{{\scriptsize{}Eclipse Platform project - Number of registers}}
\label{tab:EclipsePlatformProjectNumberRegisters}
\end{table}

\subsection{Technical and personality groups}

To find technical and personality groups of data objects that share
similar characteristics, we conducted cluster analysis through spectral
clustering. The algorithm receives as a parameter the number of clusters
($k$) in order to partition a dataset. We look for this parameter
through the elbow curve by plotting the result of the within-cluster
sum of squared errors (SSE) for different values of $k$.

Looking at the point at which the SSE value changes significantly,
we selected $k_{t}\,=\,5$ for technical clustering and $k_{p}\,=\,3$
for personality clustering. Just to clarify, a technical clustering
corresponds to the result of applying the clustering algorithm to
the data representing the files a committer has touched (i.e. a file
modified by a committer and sent by him to the repository). Personality
clustering refers to the result of applying the clustering algorithm
to the data representing personality traits inferred from committers'
texts (emails sent by committers to mailing lists).

The data representation (binary vectors representing if a committer
touched or not a file of a project) suggests that it is more convenient
to use a similarity metric like Jaccard similarity coefficient (used
in this work) than Euclidean distance.

Tables \ref{tab:ResultsTechnicalClustering} and \ref{tab:AverageNumberTimesProjectDirectoriessTouchedByCommitters}
show the results for technical clustering. The projects touched by
committers in technical clusters were obtained averaging the number
of times that project directories have been touched by the committers
in each cluster. Only values that represent a participation or contribution
of the committers to the project greater than or equal to 7\% are
taken into account.

\begin{table}[h]
\begin{centering}
\begin{tabular}{|>{\centering}m{2.5cm}|>{\centering}m{3.5cm}|}
\hline 
\textbf{\scriptsize{}Technical cluster} & \textbf{\scriptsize{}Number of committers}\tabularnewline
\hline 
\hline 
{\scriptsize{}0} & {\scriptsize{}107}\tabularnewline
\hline 
{\scriptsize{}1} & {\scriptsize{}11}\tabularnewline
\hline 
{\scriptsize{}2} & {\scriptsize{}12}\tabularnewline
\hline 
{\scriptsize{}3} & {\scriptsize{}24}\tabularnewline
\hline 
{\scriptsize{}4} & {\scriptsize{}13}\tabularnewline
\hline 
\textbf{\scriptsize{}Total} & {\scriptsize{}168}\tabularnewline
\hline 
\end{tabular}
\par\end{centering}

\centering{}\protect\caption{{\scriptsize{}Results of technical clustering.}\label{tab:ResultsTechnicalClustering}}
\end{table}

\begin{table}[h]
\begin{centering}
\begin{tabular}{|>{\centering}m{0.8cm}|>{\centering}m{0.8cm}|>{\centering}m{0.8cm}|>{\centering}m{0.8cm}|>{\centering}m{0.8cm}|>{\centering}m{0.8cm}|>{\centering}m{0.8cm}|}
\hline 
\begin{turn}{90}
\textbf{\scriptsize{}{~\parbox{1.7cm}{%
Technical cluster
}~}}
\end{turn} & \begin{turn}{90}
\textbf{\scriptsize{}{~\parbox{1.7cm}{%
Eclipse Platform
}~}}
\end{turn} & \begin{turn}{90}
\textbf{\scriptsize{}{~\parbox{1.7cm}{%
Eclipse Platform Runtime
}~}}
\end{turn} & \begin{turn}{90}
\textbf{\scriptsize{}{~\parbox{1.7cm}{%
Eclipse Platform
SWT
}~}}
\end{turn} & \begin{turn}{90}
\textbf{\scriptsize{}{~\parbox{1.7cm}{%
Eclipse Platform Team
}~}}
\end{turn} & \begin{turn}{90}
\textbf{\scriptsize{}{~\parbox{1.7cm}{%
Eclipse Platform Text
}~}}
\end{turn} & \begin{turn}{90}
\textbf{\scriptsize{}{~\parbox{1.6cm}{%
Eclipse Platform UI
}~}}
\end{turn}\tabularnewline
\hline 
\hline 
{\scriptsize{}0} & {\scriptsize{}{*}} & {\scriptsize{}{*}} & {\scriptsize{}0.13} & {\scriptsize{}{*}} & {\scriptsize{}0.07} & \textbf{\scriptsize{}0.67}\tabularnewline
\hline 
{\scriptsize{}1} & \textbf{\scriptsize{}0.24} & {\scriptsize{}0.09} & {\scriptsize{}{*}} & {\scriptsize{}{*}} & {\scriptsize{}{*}} & \textbf{\scriptsize{}0.56}\tabularnewline
\hline 
{\scriptsize{}2} & {\scriptsize{}{*}} & {\scriptsize{}{*}} & \textbf{\scriptsize{}0.73} & {\scriptsize{}{*}} & {\scriptsize{}{*}} & {\scriptsize{}0.19}\tabularnewline
\hline 
{\scriptsize{}3} & {\scriptsize{}0.07} & {\scriptsize{}0.07} & {\scriptsize{}0.08} & \textbf{\scriptsize{}0.27} & {\scriptsize{}0.12} & \textbf{\scriptsize{}0.39}\tabularnewline
\hline 
{\scriptsize{}4} & {\scriptsize{}{*}} & {\scriptsize{}{*}} & {\scriptsize{}{*}} & {\scriptsize{}{*}} & {\scriptsize{}{*}} & \textbf{\scriptsize{}0.84}\tabularnewline
\hline 
\end{tabular}
\par\end{centering}

\centering{}\begin{threeparttable}\begin{tablenotes} \footnotesize \item [*] $\mathit{Values < 0.07}$ \end{tablenotes}\end{threeparttable}\protect\caption{{\scriptsize{}Results for technical clustering. Averaged number of
times that project directories have been touched by committers.}}
\label{tab:AverageNumberTimesProjectDirectoriessTouchedByCommitters}
\end{table}

In addition, we calculate, for each technical cluster, the number
of committers who touched each project, as shown in Table \ref{tab:NumberCommittersTouchingProjectsTechnicalClusters}.
Hence, to understand the meaning of technical groups we consider the
results presented in Tables \ref{tab:AverageNumberTimesProjectDirectoriessTouchedByCommitters}
and \ref{tab:NumberCommittersTouchingProjectsTechnicalClusters}.
We noticed that most committers from all technical clusters, except
for technical cluster 2, contribute to Eclipse Platform UI project.
In fact, there is great participation of technical cluster 0 (83 committers)
and a high activity of the technical cluster 4 (0.84) with reference
to this project. Analyzing participation in other projects, we observed
that committers from cluster 0 tend to be more present in the Eclipse
Platform SWT project (35) just as committers from cluster 2 (12),
while committers from cluster 1 lean toward Eclipse Platform Runtime
(10), and committers from cluster 4 tend to work in Eclipse Platform
Team project (24). Furthermore, we noted uniformity in cluster 4 as
all the committers belonging to this group (13) contribute to Eclipse
Platform SWT, Eclipse Platform Team, Eclipse Platform Text and Eclipse
Platform UI projects, with more activity in the latter project (0.84).

\begin{table}[h]
\begin{centering}
\begin{tabular}{|>{\centering}m{0.8cm}|>{\centering}m{0.8cm}|>{\centering}m{0.8cm}|>{\centering}m{0.8cm}|>{\centering}m{0.8cm}|>{\centering}m{0.8cm}|>{\centering}m{0.8cm}|}
\hline 
\begin{turn}{90}
\textbf{\scriptsize{}{~\parbox{1.7cm}{%
Technical cluster
}~}}
\end{turn} & \begin{turn}{90}
\textbf{\scriptsize{}{~\parbox{1.7cm}{%
Eclipse Platform
}~}}
\end{turn} & \begin{turn}{90}
\textbf{\scriptsize{}{~\parbox{1.7cm}{%
Eclipse Platform Runtime
}~}}
\end{turn} & \begin{turn}{90}
\textbf{\scriptsize{}{~\parbox{1.7cm}{%
Eclipse Platform SWT
}~}}
\end{turn} & \begin{turn}{90}
\textbf{\scriptsize{}{~\parbox{1.7cm}{%
Eclipse Platform Team
}~}}
\end{turn} & \begin{turn}{90}
\textbf{\scriptsize{}{~\parbox{1.7cm}{%
Eclipse Platform Text
}~}}
\end{turn} & \begin{turn}{90}
\textbf{\scriptsize{}{~\parbox{1.6cm}{%
Eclipse Platform UI
}~}}
\end{turn}\tabularnewline
\hline 
\hline 
{\scriptsize{}0} & {\scriptsize{}11} & {\scriptsize{}17} & \textbf{\scriptsize{}35} & {\scriptsize{}22} & {\scriptsize{}19} & \textbf{\scriptsize{}83}\tabularnewline
\hline 
{\scriptsize{}1} & {\scriptsize{}7} & \textbf{\scriptsize{}10} & {\scriptsize{}6} & {\scriptsize{}5} & {\scriptsize{}7} & \textbf{\scriptsize{}11}\tabularnewline
\hline 
{\scriptsize{}2} & {\scriptsize{}2} & {\scriptsize{}0} & \textbf{\scriptsize{}12} & {\scriptsize{}7} & {\scriptsize{}11} & \textbf{\scriptsize{}12}\tabularnewline
\hline 
{\scriptsize{}3} & {\scriptsize{}13} & {\scriptsize{}12} & {\scriptsize{}13} & \textbf{\scriptsize{}24} & {\scriptsize{}14} & \textbf{\scriptsize{}24}\tabularnewline
\hline 
{\scriptsize{}4} & {\scriptsize{}9} & {\scriptsize{}6} & \textbf{\scriptsize{}13} & \textbf{\scriptsize{}13} & \textbf{\scriptsize{}13} & \textbf{\scriptsize{}13}\tabularnewline
\hline 
\end{tabular}
\par\end{centering}

\centering{}\protect\caption{{\scriptsize{}Number of committers touching projects in technical
clusters.}\label{tab:NumberCommittersTouchingProjectsTechnicalClusters}}
\end{table}

\begin{figure}[h]
\begin{centering}
\includegraphics[width=0.7\columnwidth]{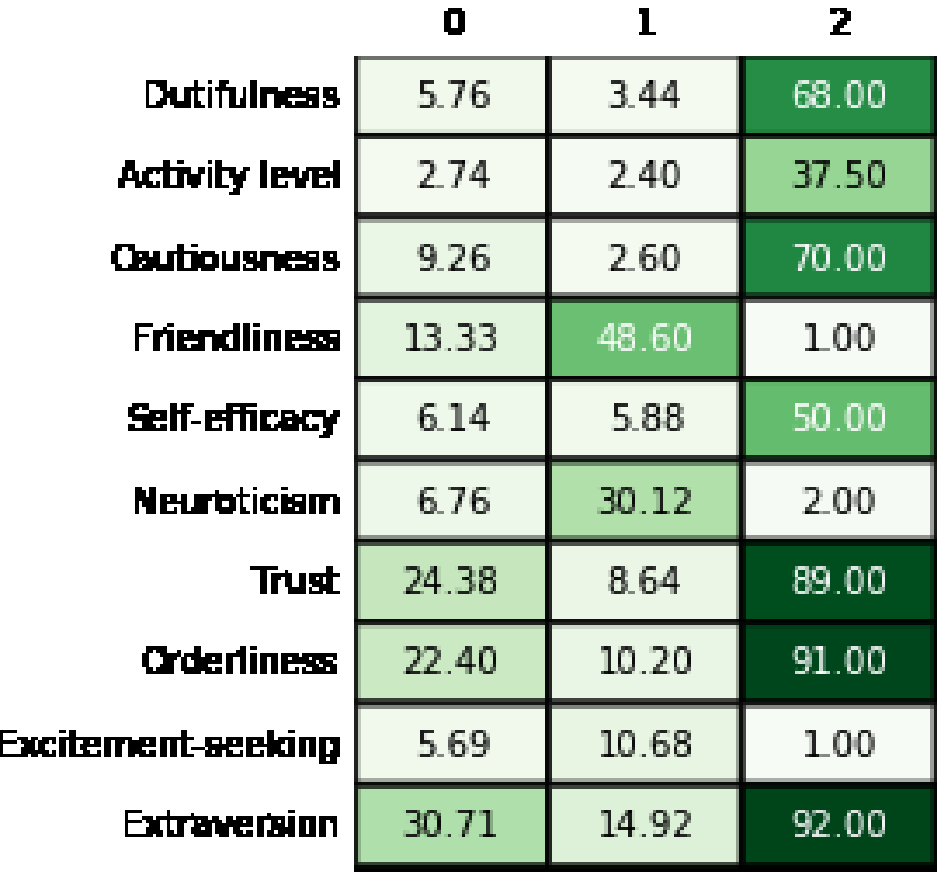}
\par\end{centering}

\protect\caption{{\scriptsize{}Heat map of the most discriminative factors for the
personality clustering.}}

\centering{}\label{fig:heatmapPersonalityInsightsClusteringResults}
\end{figure}

Table \ref{tab:ResultsPersonalityClustering} shows the results for
personality clustering, and the heat map in Figure \ref{fig:heatmapPersonalityInsightsClusteringResults}
depicts the results of each Big Five dimension and facet, each Need,
and each Value (rows) by each personality cluster (columns). Because
of space restrictions, we show just the top-10 (the lowest) entropy
values for Big Five dimensions and facets, needs, and values. As recommended
by the IBM Watson Personality Insights service and for statistically
significant results, we analyzed at least 3,500 words written by each
committer. To get enough text for each committer, we concatenated
his/her e-mails sent to the project mailing lists.

\begin{table}[h]
\begin{centering}
\begin{tabular}{|>{\centering}m{2.7cm}|>{\centering}m{3.5cm}|}
\hline 
\textbf{\scriptsize{}Personality cluster} & \textbf{\scriptsize{}Number of committers}\tabularnewline
\hline 
\hline 
{\scriptsize{}0} & {\scriptsize{}42}\tabularnewline
\hline 
{\scriptsize{}1} & {\scriptsize{}24}\tabularnewline
\hline 
{\scriptsize{}2} & {\scriptsize{}2}\tabularnewline
\hline 
\textbf{\scriptsize{}Total} & {\scriptsize{}68}\tabularnewline
\hline 
\end{tabular}
\par\end{centering}

\centering{}\protect\caption{{\scriptsize{}Results of personality clustering.}\label{tab:ResultsPersonalityClustering}}
\end{table}

Then, as answer to RQ1, the personality traits can be identified through
communications between software developers involved in FLOSS projects,
which are those corresponding to the Big Five dimensions and facets,
needs, and values%
\footnote{{\tiny{}\url{www.ibm.com/smarterplanet/us/en/ibmwatson/developercloud/doc/personality-insights/models.shtml}}%
}. As highlighted in the heat map, personality cluster 2 groups the
committers with the highest scores in personality traits such as Extraversion
(92\%), Orderliness (91\%), Trust (89\%), Cautiousness (70\%) and
Dutifulness (68\%), and the lowest values in Excitement-seeking (1\%),
Friendliness (1\%) and Neuroticism (2\%). Personality traits characterizing
cluster 1 by its moderately high values are Friendliness (48.6\%)
and Neuroticism (30.12\%), opposed to cluster 3 which has very low
values in those personality dimensions. Finally, personality traits
standing out in cluster 0 are Trust (24,38\%), Orderliness (22.40\%)
and Extraversion (30.71\%), which are lower than those of cluster
2, but higher than those of cluster 1.

\subsection{Personality traits characterizing technical groups}

Each personality cluster has associated personality traits that characterize
and distinguish its members. From the results of the IBM Watson Personality
Insights service it is possible to identify which personality traits
are dominant in each cluster, becoming differentiating features, and
what personality traits have similar values across all groups. In
addition, we know which technical group is associated to each committer.

By computing the entropy for each of the Big Five dimensions and facets,
Need and Values, it is possible to determine which of these attributes
provide more information or become a differentiating factor when analyzing
the technical groups, depending on the personality traits of the committers
who are part of them. The lower the entropy, the greater the variation
of the values of the corresponding attribute for the technical clusters,
i.e., the attribute turns out more informative. This allows us to
characterize the group or groups in which it is presented, and in
which we must focus on when making an analysis of each cluster.

\begin{figure}[h]
\begin{centering}
\includegraphics[width=1\columnwidth]{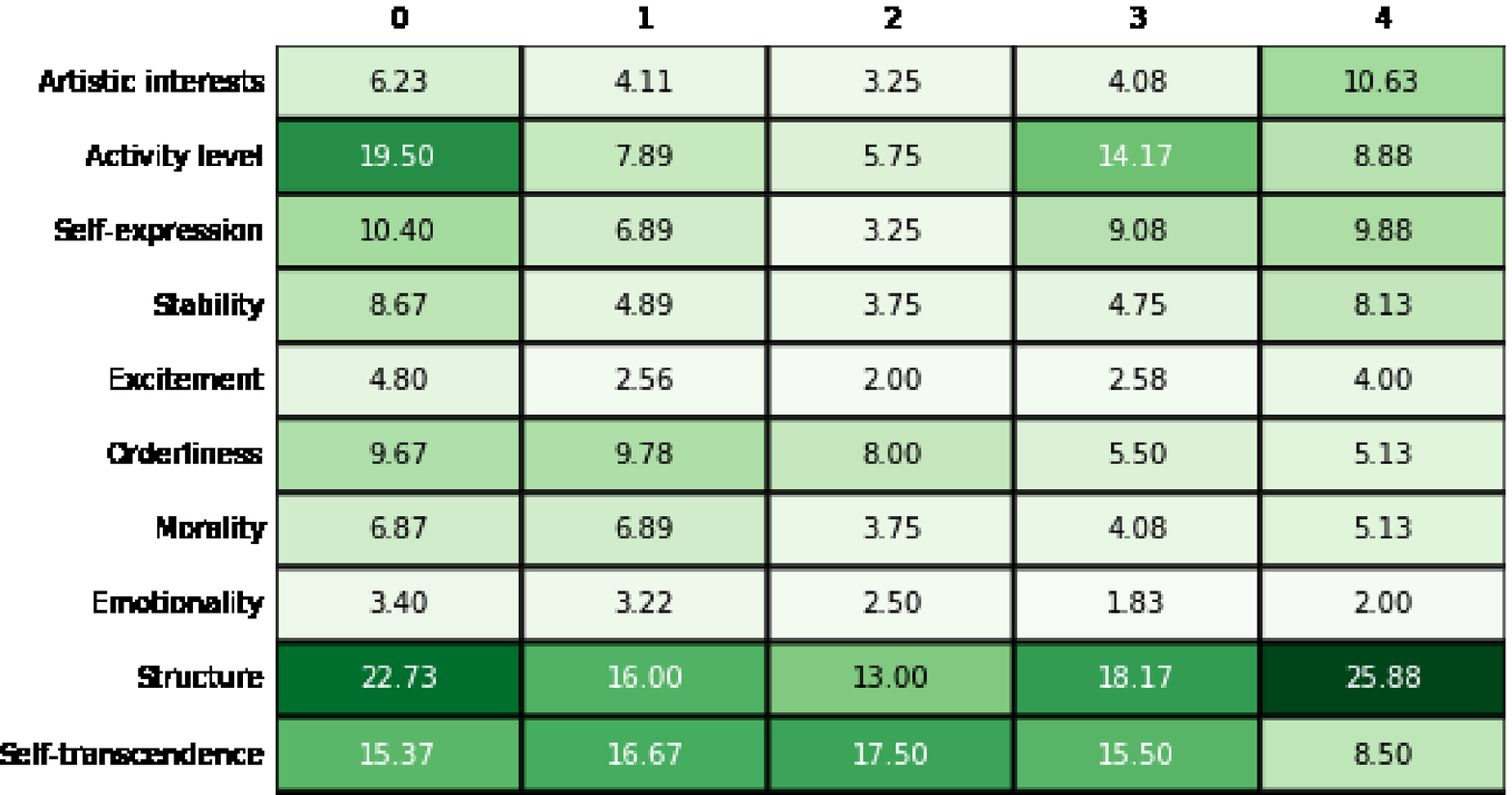}
\par\end{centering}

\protect\caption{{\scriptsize{}Heat map of personality centroids for each technical
cluster.}}

\centering{}\label{fig:heatmapPersonalityCentroidsForEachTechnicalCluster}
\end{figure}

Since we know the technical cluster where each committer belongs and
the personality traits for committers belonging to each technical
cluster, we can compute the centroids of personality traits for each
technical cluster. Again, due to space restriction, we show just the
10 lowest values and the 10 highest values of entropy for the Big
Five dimensions and facets, Needs, and Values of personality centroids
computed by averaging the values of the personality traits of committers
in each technical cluster. Figure \ref{fig:heatmapPersonalityCentroidsForEachTechnicalCluster}
shows the results.

From the results reported in Figure \ref{fig:heatmapPersonalityCentroidsForEachTechnicalCluster}
and Table \ref{tab:AverageNumberTimesProjectDirectoriessTouchedByCommitters},
we can answer RQ2. Personality traits scoring high ($\geqslant80\%$)
and with nearly uniform values through all technical clusters (e.g.
\textit{Cooperation}, \textit{Sympathy}, \textit{Conscientiousness},
\textit{Achievement striving}, \textit{Cautiousness}, \textit{Openness},
\textit{Adventurousness}, \textit{Imagination}, \textit{Intellect},
\textit{Liberalism}, \textit{Conservation} and \textit{Self-enhancement})
could be considered as personality factors characterizing the project,
i.e. people involved in the project will most likely exhibit high
values in these personality traits. On the other hand, personality
traits scoring lower ($<25\%$) allow us to identify relationships
with the technical aspects, differentiating personality features among
the different technical clusters.

\begin{figure*}[t]
\begin{centering}
\includegraphics[width=2\columnwidth,height=2\columnwidth,keepaspectratio]{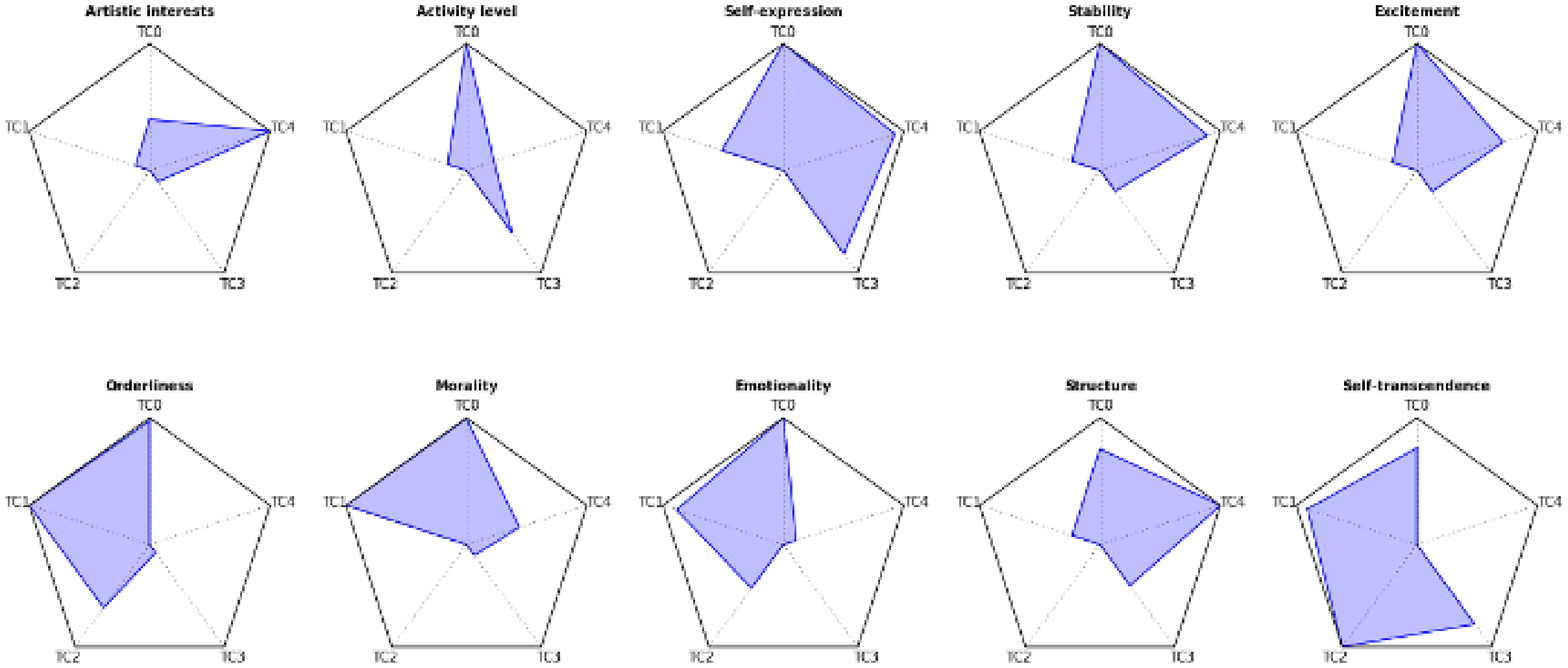}
\par\end{centering}

\protect\caption{{\scriptsize{}Radar charts of personality traits by technical cluster
(TC).}\label{fig:RadarChartsPersonalityTraitsTechnicalClusters}}
\end{figure*}

Figure \ref{fig:RadarChartsPersonalityTraitsTechnicalClusters} summarizes
personality traits by technical cluster allowing to visualize which
personality traits are more representative in each technical cluster.
From this representation, one can notice the dominant facets for the
different technical clusters. For instance, committers grouped in
the technical cluster 4 score high values in the \textit{Artistic
interests} facet in comparison with other clusters, and they mainly
contribute to a project related to graphical elements, i.e., Eclipse
Platform UI. Furthermore, a high value in \textit{Structure} need%
\footnote{{\tiny{}\url{https://www.ibm.com/smarterplanet/us/en/ibmwatson/developercloud/doc/personality-insights/models.shtml\#outputNeeds}}%
} (25.88\%), and a low value in \textit{Self-transcendence} value%
\footnote{{\tiny{}\url{https://www.ibm.com/smarterplanet/us/en/ibmwatson/developercloud/doc/personality-insights/models.shtml\#outputValues}}%
} (8.5\%) regarding the other clusters could explain why the committers
of the technical cluster 4 contribute to only one project.

\subsection{Visualizing the social network - from committers to mailing lists}

\begin{figure*}[t]
\begin{centering}
\includegraphics[width=1.1\columnwidth,height=1.1\columnwidth,keepaspectratio]{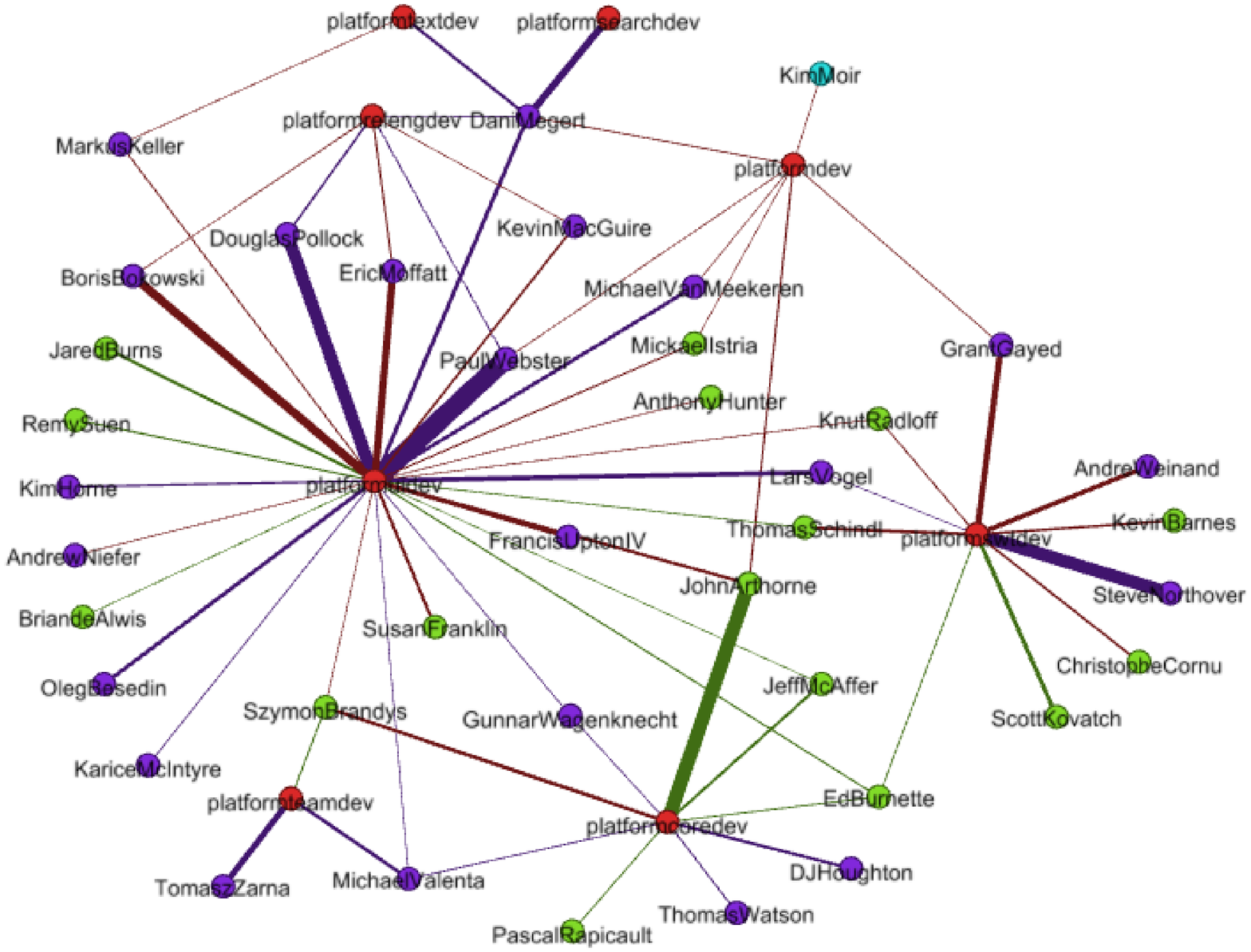}
\par\end{centering}

\protect\caption{{\scriptsize{}Social (email communication) network. From committers
to mailing lists.}}

\centering{}\label{fig:emailCommunicationNetworkPersonalityClusters}
\end{figure*}

Using the e-mails sent by committers to the Eclipse Platform project
mailing lists, we built a graph representing e-mail communications.
The graph in Figure \ref{fig:emailCommunicationNetworkPersonalityClusters}
shows committers and mailing lists (PlatformDev, Search, Text, Core,
Releng, UI, SWT, Team, i.e., red circles) as nodes. The thickness
of the edge between a committer and a mailing list represents the
amount of emails sent by the committer to the list. Additionally,
the color of the nodes representing committers corresponds to the
personality cluster to which the committer belongs to. Only committers
that have sent more than 10 e-mails to any of the lists were taken
into account.

Figure \ref{fig:emailCommunicationNetworkPersonalityClusters} helps
us to answer RQ3. Committers belonging to personality group 0 (purple
circles) are those distributed through all mailing lists, except Team.
This may be attributed to the ranking they have in traits such as
Altruism (73.14\%), Cheerfulness (70.93\%), Gregariousness (89.05\%),
and Self-discipline (46.62\%). The top 5 of committers, sorted by
the number of messages sent to mailing lists, is distributed between
representatives of the personality clusters 0 and 1. This tendency
to actively participate in the lists may be related to the traits
having similar values in those clusters such as \textit{Gregariousness}
(89.05\% for cluster 0 and 76.88\% for cluster 1), and \textit{Altruism}
(73.14\% for cluster 0 and 63.04\% for cluster 1). Only one committer
appears in the graph of Figure \ref{fig:emailCommunicationNetworkPersonalityClusters}
representing the personality group 2, which has the highest value
of \textit{Cautiousness} (9.26\% for cluster 0, 2.60\% for cluster
1, and 70\% for cluster 2) and the lowest value of \textit{Cautiousness}
(1\%), compared to the other groups (13.33\% for cluster 0 and 48.60
for cluster 1); but surprisingly this group has the highest value
of \textit{Extraversion} (30.71\% for cluster 0, 14.92\% for cluster
1, and 92\% for cluster 2) and for \textit{Dutifulness} (5.76\% for
cluster 0, 3.44\% for cluster 1, and 68\% for cluster 2).

\section{Conclusions}

FLOSS projects are characterized by a high component of social interaction
where a large number of people with great technical skills contribute
from different parts of the world, in most cases without knowing each
other. Within this context we conducted a preliminary study aimed
at uncovering the factors involved in the formation of working groups
and the dynamics of communication that occur during the process of
software development.

Considering that personality traits influence most, if not all, of
the human activities, we took this feature as the centerpiece of the
work done. In this regard, services such as IBM Watson Personality
Insights are crucial to analyze personality traits from text, when
is impractical to apply personality tests to each participant of a
study. By having the personality characteristics (a total of 52) inferred
by the service, we were able to identify relationships established,
either solely from personality traits as is the case of the groups
presented in Table \ref{tab:ResultsPersonalityClustering}, or those
established from both personality traits and social activities of
the committers related to communication through the project mailing
lists, as shown in Figure \ref{fig:emailCommunicationNetworkPersonalityClusters}.

As evidenced by analyzing the graph representing social activities
(Figure \ref{fig:emailCommunicationNetworkPersonalityClusters}),
is not enough to focus on just one personality trait to identify patterns
due to the complexity of the personality and its constitutive factors.
Thus, it is necessary to give a comprehensive and detailed look at
each one of the dimensions, facets and categories of the three personality
models (Big Five, Needs and Values) to be able to draw conclusions
most closely related to the behavior the data try to show us.

\section{Threats to validity}

What we must have in mind is that the main aim of this work is to
explore whether it is possible to extract personality traits from
developer e-mails, and try to uncover relationships among those traits
and the social and technical activities performed by the software
team. As a feasible way to achieve this goal, we proposed a novel
approach to collect, process, and analyze the relevant data, which
involves the use of several tools and clustering techniques.

We are aware that our preliminary results may be affected by several
validity threats inherent in the proposed approach. To mention only
the most important ones, our results depend on an automatic analysis
of developer e-mails performed by IBM Watson Personality Insights
service, instead of personality assessment questionnaires designed
by psychologists and applied directly to the software team members.
Moreover, our experiment is limited to the mailing lists and code
base of one system only. Thus, variables such as the project domain,
the system size, the team size, and the quality and availability of
the text could influence the effectiveness of our approach.

\end{document}